\documentclass[conference]{IEEEtran}
\newcommand{\relsim}{AIReSim\xspace}

\usepackage{xspace}
\usepackage{graphicx} 
\usepackage[colorinlistoftodos,textsize=scriptsize]{todonotes}
\usepackage{enumitem}
\usepackage{listings}
\usepackage{multirow}
\usepackage{amssymb}
\usepackage{pifont}
\usepackage{algorithm}
\usepackage{algpseudocode}
\usepackage{hyperref}

\usepackage{tikz}
\usepackage{amsmath}
\usepackage{enumitem}
\usepackage{caption}
\usepackage{listings}
\usepackage{xcolor}

\newcommand{\rev}[1]{{#1}}
\usepackage[table,xcdraw]{xcolor}
\usepackage{tabularx}

\title{\relsim: A Discrete Event Simulator for Large-scale AI Cluster Reliability Modeling}
\author{
    \IEEEauthorblockN{Karthik Pattabiraman\IEEEauthorrefmark{1}, 
    Mihir Patel\IEEEauthorrefmark{2}, and 
    Fred Lin\IEEEauthorrefmark{2}}
    
    \IEEEauthorblockA{\IEEEauthorrefmark{1}University of British Columbia (UBC), karthikp@ece.ubc.ca}
    
    \IEEEauthorblockA{\IEEEauthorrefmark{2}Meta Platforms, \{patelmc, fanlin\}@meta.com}
}

\begin{document}

\maketitle

\begin{abstract}
Failures in clusters running large-scale AI workloads can result in decreased utilization. 
Because the cost of a failure in such AI workloads is high (as it requires restarting the entire job from a previous checkpoint), there are many mechanisms in place to ensure that the failures are mitigated, and the impact of a failure is minimized. However, these mechanisms have many knobs and parameters, all of which must be carefully tuned based on the system and cluster's characteristics. We built 
\relsim, a discrete event simulator to evaluate the different design choices during the failure, recovery, scheduling and repair processes for a cluster running a large-scale AI workload. 
\relsim allows the system designer to systematically evaluate the effects of the different knobs and parameters on the overall end-to-end reliability of the system. Further, \relsim can be used to identify which knobs or parameters are important in order to prioritize the investment of effort in improving the system. \relsim also allows tuning of the knobs for achieving different tradeoffs in the system, as well as to consider various ``what-if'' scenarios. We present a case study of applying \relsim for capacity planning for large-scale clusters running AI workloads. 
\end{abstract}

\section{Introduction}
\label{Sec:intro}

Artificial Intelligence (AI) and Machine Learning (ML) applications have become ubiquitous in many areas, especially with the advent of generative AI such as ChatGPT~\cite{chatgpt2025}. These applications require enormous computational resources for training, which necessitate the use of large computational clusters numbering thousands of high-performance servers equipped with Graphic Processing Units (GPUs) and other specialized accelerators~\cite{mishra2023artificial}. Unfortunately, at these scales, hardware failures become commonplace, both due to random and systematic causes~\cite{kokolis2025revisiting}. For example, 
Meta reported a total of 466 failures during a 54 day period when training their Llama3  models, of which 78\% were due to hardware~\cite{grattafiori2024llama}. These frequent failures necessitate costly recovery techniques such as checkpointing, thereby negatively affecting the throughput of the AI training job, and resulting in a net loss of utilization (e.g., OpenAI reports an average GPU utilization of only 32-36\% when training the GPT4 job~\cite{Cunchi-2025}). This is because if even a single server\footnote{We refer to individual machines in the cluster as servers.}  fails during the training, the entire job needs to be restarted (typically from a previous checkpoint). 

There are two kinds of failures typically encountered by training jobs. The first are random failures, which as the name suggests, occur at random and are not typically reproducible, e.g., a bit flip due to cosmic rays that corrupts a value in memory, thereby causing the server to crash. The second are systematic failures, which occur due to 
 underlying issues such as manufacturing variations in hardware, temperature fluctuations, or due to rare software defects and configuration errors. These failures typically manifest in the same servers over a period of time (however, they are not necessarily deterministic in that they may manifest at random times). These failures are often a more serious concern than random failures, as they can result in the same servers crashing repeatedly. Further, recent studies have shown that the rate of systematic failures is orders of magnitude higher than random failures~\cite{dixit2021silent, Mitra-2025}. Therefore, it is important to mitigate systematic failures in order to boost the overall reliability and throughput of the training job.

 The common way to mitigate systematic failures is to perform repair of all failed servers, by taking them offline for testing and potential repair after a failure. 
 However, doing so naively can be expensive, and reduce the throughput of the AI training job, as it needs a certain minimum number of servers to meaningfully execute (e.g., 4096 servers). If there are fewer servers, then the job will be slow or may even stall. Therefore, we need to provision a sufficient number of servers as spares  when a server is sent for repair, to continue the job. However, allocating too many servers as spares is wasteful as they consume energy and take away resources from other jobs. 

 The above is one example of a tradeoff that arises in clusters used for large-scale AI training jobs. There are many knobs and parameters involved in configuring these clusters, and it is important to ensure that  these knobs are calibrated and configured for optimal reliability and throughput. Traditionally, there are two possible approaches to this problem, (1) analytical (mathematical models) methods such as Markov models~\cite{trivedi2001probability}, and (2) discrete event simulation (DES)~\cite{fishman2001discrete}. The former is much faster but suffers from simplistic assumptions (e.g., Markovian arrivals), which affects its accuracy. Therefore, DES is often used to model more detailed and realistic situations, especially when failure arrival rates do not correspond to standard  distributions~\cite{fishman2001discrete}. 

 In this paper, we propose \relsim (AI Reliability simulator - pronounced airysim), a DES for simulating the reliability of large-scale clusters for running AI training jobs. \relsim is built specifically for modeling commonly used failure handling and recovery mechanisms in AI clusters, and is easily extensible to model other mechanisms. 
 
 We use \relsim to determine the values of different parameters and knobs in the cluster, e.g., choose the optimal number of spares needed for tolerating failures (for a given failure rate and recovery time). We also use \relsim to perform a parameter sweep over the set of knobs to understand the sensitivity (importance) of different parameters. Finally, \relsim can be used to perform experiments on ``what-if" scenarios and evaluate the results (e.g., what if the failure rate was to be different in the future). 

 In the rest of this paper, we first provide a general background on AI clusters and their reliability (Section~\ref{sec:background}), and then describe the standard assumptions and failure-recovery mechanisms in AI cluster. We then describe the architecture of \relsim and its features (Section~\ref{sec:approach}). Finally, we present the results of our experiments on doing a parameter sweep with \relsim  (Section~\ref{sec:evaluation})\footnote{The parameter values shown are illustrative and do not reflect the actual cluster settings; they are chosen to represent the tradeoffs in the real cluster.} for capacity planning,  before concluding the paper (Section~\ref{sec:conclusion}). 

\section{System Model and Background}
\label{sec:background}

We first present an overview of the faults and failures that can occur in large-scale AI training cluster, along with a description of the main challenges in dealing with them. We then provide an overview of the standard failure mitigation mechanisms that are often deployed in these clusters. Finally, we provide a brief background on DES.

\subsection{Faults and Failures in AI Clusters}

An AI cluster consists of a large number of components, processors, GPUs, memories, network switches, routers etc. Each of these components can fail during deployment. 
As mentioned, there are two dominant kinds of failures that occur in these clusters (or any other large-scale cluster), namely (1) random failures, and (2) systematic failures. Random failures are caused by events such as cosmic rays, temperature fluctuations and very rare software errors. Systematic failures, on the other hand, are caused by repetitive events such as manufacturing issues, wear and tear due to aging, software configuration issues, etc. Typically, hardware failures follow a classic ``bath-tub'' curve, with most of the systematic issues manifesting at both ends of the curve~\cite{trivedi2001probability}, while the flat portion of the curve (operational phase) consists mostly of random failures. However, modern hardware is becoming increasingly difficult to test thoroughly during the initial phase of the bathtub curve, due to increased variability and complexity. As such, systematic failures are becoming more common even during the operational phase and thus we need a way to deal with these systematic failures throughout the lifecycle (including the operational phase) of the cluster ~\cite{dixit2022detectingsilentdatacorruptions}.

Modern AI training jobs (we call a single execution of a workload as a job), typically consist of multiple tasks running across many thousands of nodes, each equipped with high performance processors and GPUs. These jobs  typically execute in task synchronous parallelism mode. In other words, each task communicates periodically with other tasks in an asynchronous manner e.g., exchanging gradients. The failure of a single task results in the failure of the entire job. This is because the tasks need to exchange information periodically with each other, and the loss of the information gathered by a task can cause the other tasks' data to become invalid. Thus, they all need to be restarted upon a failure. 

To deal with failures, most AI workloads use checkpointing and recovery, a classic technique used in the high-performance computing (HPC) domain~\cite{elnozahy2004checkpointing}. Each task periodically takes a (asynchronous) checkpoint, writing its state to secondary storage or persistent memory that persists even after the task fails. When a task fails, the checkpoint is loaded into a new task on a different server,  and the entire job is restarted from the checkpoint. Because checkpoints are taken asynchronously, they incur minimal overheads on the execution of the job, and are hence taken frequently. However, the process of loading the checkpoint and restarting the job after a failure on a different server incurs a significant latency (i.e., recovery time). 

Therefore, every failure incurs a performance penalty on the job, and hence it is better to avoid failures if possible. 
While random failures cannot be avoided (usually), it is possible to avoid systematic failures by not using servers that are prone to exhibiting such failures. Prior work~\cite{jiao-2025} has shown that a small proportion of the servers exhibit highly elevated rates of systematic failures, and are hence responsible for the major portion of such failures. Unfortunately, it is not possible to identify such servers apriori and remove them from the cluster~\cite{Jiang-2025}. We describe strategies to mitigate the effects of these failures in the next section.

\subsection{Failure Mitigation Mechanisms}

When a server (node) fails, the task running on that server also fails, and so does the entire AI job (as explained in the previous section). This means that if we somehow exclude nodes that are highly prone to (systematic) failures, we can reduce the total number of failures, and hence the consequent performance penalty of failures. The challenge is in identifying such nodes. As mentioned, unlike the classical bathtub curve in which systematic failures predominantly occurred in the infant mortality stage or the wear out phase, systematic failures today occur even during the normal operational phase. However, not all failures are systematic, and therefore, we cannot assume that when a node fails, it is due to a systematic failure. Therefore, we need to run tests (and repairs) on a failed node to identify the kind of failure it experienced, and either reintegrate it into the cluster if it is a random failure or the systematic failure was repaired, or to remove it from the cluster if we determine it is a systematic failure that cannot be repaired (note that we can alternatively remove it only after it experiences a certain number of systematic failures if we want to be conservative) - \rev{this is known as server retirement.}

In general, there are two kinds of tests and repairs that are possible. First, there is automated testing and repair, which aims to perform completely automated testing for common issues and attempt to resolve them without human intervention. While limited in scope, these repairs are much faster than manual repairs, and do not require costly human labour. The second kind, manual repair, has a more extensive scope, but requires costly human labour to look into the problem and then repair it. Consequently, manual repair often takes much longer than automated repair, but has a lower failure rate. Therefore, most servers first undergo automated repair, and then if that fails to fix the problem, they get sent to manual repair. Note however that both kinds of repairs can fail, and the server may be reintegrated into the cluster even if the (systematic) failure  has not been successfully repaired. It is also possible to falsely determine a server has a systematic failure, and remove it unnecessarily from the cluster (thus reducing the cluster's capacity) ~\cite{hwremediationatscale}. 

In case a repair is deemed successful (either manual or automated), we may reintegrate a server into the cluster even if the problem persists. In this case, the server may experience another (systematic) failure, and go through the entire process again. To alleviate the effects of such repeated failures, we may maintain a score for each server that keeps track of how often it has failed in a given time period, and remove servers that exhibit a number of failures exceeding a certain threshold (within that time period).

When a server is sent to repair, either automated or manual, it becomes unavailable for running of tasks during that time. Further, if a server is removed permanently from the cluster, it is no longer available for running tasks. Therefore, there needs to be spare capacity in the cluster to deal with these temporary and permanent server unavailability. This is typically done in the form of having spare servers that are part of the ``working pool'' of servers that are ready to take over when a server become unavailable. Because these servers have to be ready to take over at short notice, they cannot be used to run other workloads, and have to be powered on (thus consuming energy and taking up cluster resources). In contrast, we have servers in a separate pool called the ``spare pool'', that are used to run other jobs (not AI training jobs though). When these servers need to be used for the AI training job, there is a delay to preempt the other jobs and provision the server for use in the AI job. This typically happens only when we run out of servers in the working pool. Because spare pool servers are used to run other jobs, they are not wasteful in terms of their energy and resource consumption. When the need for additional servers for the AI job subsides, these servers are returned to the spare pool. In the hypothetical event we run out of servers in both the working and spare pools, the job stalls until servers come back from repair. This is a situation to be avoided, as it can result in significant delays in job completion times. 

Finally, there is a scheduler that decides at the start of each job iteration (at the start of the job and after each failure), which servers to allocate for the job - this process is also known as host selection. The end to end time including host selection and job (re)start is non-trivial, and hence it helps to minimize the number of times that a job gets in this state. One way to do this is, each job is started with an additional number of servers than is strictly necessary - these servers are allocated to the job, and when a failure occurs, they can be swapped in for the server that failed \emph{without} going through the host selection, job (re)start process. For example, a job that needs 4096 servers is allotted (say, 4096+16) servers, so that it can tolerate up to 16 server failures before going in for host selection again. We call these servers \emph{warm standbys}. Note that we can tolerate even more failures if servers are returned back to the pool after repair, before the job runs out of these 16 servers (a server is returned to the job after repair if it was originally assigned to the same job before it failed, without going through host selection again). 

\subsection{Discrete Event Simulation (DES)}

DES is the process of modeling a system 
and running different scenarios for the system. Unlike analytical modeling techniques~\cite{SHARPE}, simulation is more realistic as it can deal with complex distributions, and is not restricted to only considering Markovian failure distributions and independence assumptions. It is also easier to try different scenarios in simulation than on the real system in the form of A/B testing, which can be expensive and risky~\cite{quin2024b}.

There are three  uses of DES. First, we can use it to obtain point estimates of different configurations of the system and see if they are adequate. Second, we can use it to understand the effects of different parameter values on the system, and to prioritize our efforts. Finally, we can use it to determine the optimal sets of parameter values for the system for a given set of conditions, and also tune them.

In this paper, we build \relsim, a DES for simulating the operations of a cluster running AI workloads. We are interested in simulating the behavior of the AI training job under failures, recovery and repairs. Specifically, we are interested in understanding the effects of the different parameters (knobs) on the system, as well as the heuristics we use in the system. We are also interested in simulating the effects of future trends on the system, and studying their effects, e.g., what if failure rates increase and whether current policies will still be effective.

With \relsim, we can consider multiple parameters at the same time, and also ask questions of the following form ``what would happen if we improved this part of the pipeline or implemented this technique?''. For example, how much does the target measure (e.g., availability) improve if we reduce the recovery time after a failure by 50\%. Likewise, ``when the same server fails repeatedly, after how many failures should we remove it from the cluster for ever?'' While analytical modeling can be used to answer these questions, we would face limitations due to the kinds of assumptions it makes and the scope. \emph{In contrast, \relsim is a flexible and general-purpose infrastructure to answer questions like the above.}

\rev{General-purpose DES frameworks such as SimGrid~\cite{simgrid} and SimPy~\cite{scherfke2021simpy} can in principle model arbitrary distributed systems, but they require significant domain-specific engineering to capture AI cluster reliability constructs. \relsim provides these constructs — warm standbys, repair processes, restarts — as first-class primitives, reducing the effort needed to model and evaluate real AI cluster configurations. \emph{To the best of our knowledge, no existing open-source tool combines AI-cluster-specific reliability modeling with a pluggable policy framework and parameter sweep infrastructure, as \relsim does.}}

\section{Approach: \relsim}
\label{sec:approach}

In this section, we describe the design and implementation of \relsim. We first explain the assumptions behind the design of \relsim. We then explain the inputs and outputs of \relsim, followed by a high-level view of its structure. Finally, we present the implementation details. 

\subsection{Assumptions and Metrics}

\textbf{Assumptions}. We make seven assumptions as follows. 

\begin{enumerate}
\item We have two kinds of failures in the cluster: (1) systematic and (2) random. 
We call the servers exhibiting systematic failure with elevated failure rates as bad servers, and the servers exhibiting random failures with normal failure rates as good servers. However, we do not know apriori which server is which, though we know the fraction of bad servers in the cluster. We consider two cases, one where the bad servers are all initialized at the beginning of the simulation and never change, and the other where bad servers are regenerated periodically (e.g., end of life aging or new hardware models being integrated into the cluster).
\item Failures occur based on the exponential distribution, and the failure rates for both systematic and random failures are constant. Exponential distributions are a common assumption, and can be modeled with a single parameter - this is why we assume the same. However, \rev{\relsim also supports the Lognormal and  Weibull distributions. It can also be extended with user-specified distributions.}
\item \relsim models repairs at a high level of abstraction, and hence does not consider all the details of the repair process such as specific repair actions etc. We assume that first, upon a failure, a server is sent to automatic repair. If the problem is determined to be beyond the scope of automated repair, it is sent to manual repair (based on a probability specified in the simulation). Both kinds of repairs may or may not succeed (again based on a probability in the simulation). \item We assume that repair events are independent of each other as we are considering repairs at an abstract level without considering specific repair actions. Further, we assume there is only one repair action initiated per failure.  Note that repairs are considered as unavailable duration of the servers, and we assume that repair durations are exponentially distributed, with the specified average times. 
\item If a repair succeeds, a bad server becomes good. Otherwise, it stays bad. This is the case for both manual and automated repairs. We assume that repairs are stateless as we are considering repairs at a high level of abstraction. We can extend \relsim to consider stateful repairs. 
\item At the beginning of the simulation, we start with empty queues. Note that we assume there is only one AI job executing at any time in the cluster. However, this can be easily modified in the simulator if needed, e.g., to consider multiple concurrent AI jobs.
\item We only model a server’s failure and recovery when it is executing the (AI) job - all other jobs are not modeled in the system. \relsim supports accounting for the cost of preempting other jobs. This is done by assuming a fixed cost per server (in units of time) for each server in the job that was preempted.
\end{enumerate}


\subsection{Inputs and Outputs}

\textbf{Input Parameters}: \relsim takes the following parameters as inputs to the simulation. 

\begin{enumerate}
\item Failure rate (both random and systematic):  Failure rate of servers that have random and systematic failures respectively. Recall that we assume they are exponentially distributed; therefore, the failure rate is sufficient to characterize the distribution. 
\item Fraction of systematic failures: Fraction of servers with systematic failures in addition to random failures. 
\item Recovery time after failures: Time taken to recover the job after a failure (e.g., 10 minutes)
\item Job size (e.g., 4K job): How many servers are needed for the job (e.g., 4096)
\item Job length: How long does the AI job take to run, without failures  (e.g., 256 days)
\item Number of warm standbys for each job: How many servers are alloted as warm standbys (e.g., 32)
\item Working Pool size: How many servers are in the working pool (e.g., 4200)
\item Spare Pool Size: How many servers are in the spare pool (e.g., 500)
\item Repair times (both automated and manual): Time taken for automated and manual testing and repairs (e.g., 30 minutes and 3 days respectively)
\item Repair failure probability (both automated and manual): Probability that the repair action initiated does not resolve the issue, though the status says the repair succeeded - thus, the repair failed.
\item Probability of automated repair going to manual repair: Probability that the initial automated repair does not work and the server is sent to manual repair instead (note that this is different from a repair failure)
\item \rev{Diagnosis probability: The failure was diagnosed and a server is identified (may or may not be the correct server).}
\item Diagnosis uncertainty: The failure was diagnosed incorrectly and the wrong server was identified as the cause.
\end{enumerate}

\textbf{Outputs}: \relsim measures the following outputs from the simulation runs. (1) Total time taken to train the AI job, (2) Total number of failures, as well as numbers of random and systematic failures, (3) Number of preemptions (if any), (4) Average number of repairs (both manual and automated), and (5) Average run duration of the job. It calculates common statistics such as mean, median, standard deviation and order percentiles for each of the above outputs. 

\subsection{Design}

At a high level, there are five different aspects of the system that are captured by \relsim, namely
(1) Failures of servers, both random and systematic
(2) Recovery of the server after failure, including removing servers from the working pool, 
(3) Host selection and allocation of servers to the job, including warm standbys, and 
(4) Repair of failed servers, both manual and automated,
(5) Moving of servers from spare pool to working pool and back, as needed

\relsim consists of five modules corresponding to the above five aspects as follows. We first explain the functionality of the modules, and then the algorithm of \relsim. 
\begin{enumerate}
\item \textit{Server}: Keeps track of each server’s failure and recovery. When a job is started on a server, a failure process starts at the same time. If the failure process triggers first, the job is interrupted and the server fails. The server notifies the coordinator of its failure.  \rev{Note that we approximate this process by analytical calculation of the failure rates.}
\item \textit{Coordinator}: This module coordinates the execution of the entire group of servers. When a server fails, the coordinator is notified. In turn, it informs the other servers in the group of the failure, and asks them to stop executing the job (and initiate a fast recovery).
\item \textit{Scheduler}: Assigns servers to the job from a list of chosen servers (host selection), and starts the job on the servers. It also keeps track of the remaining length of the job and failed servers. It also implements different methods of choosing servers for the job. 
\item \textit{Repairs}: Models the repair of servers as a separate process, in parallel with the failures and computation. It is used for both manual and automated repairs. 
\item \textit{Pool}: Keeps track of the servers in working and spare pools, and moves servers between them if needed. 
\end{enumerate}

\begin{figure}
    \centering
    \includegraphics[width=1.0\linewidth]{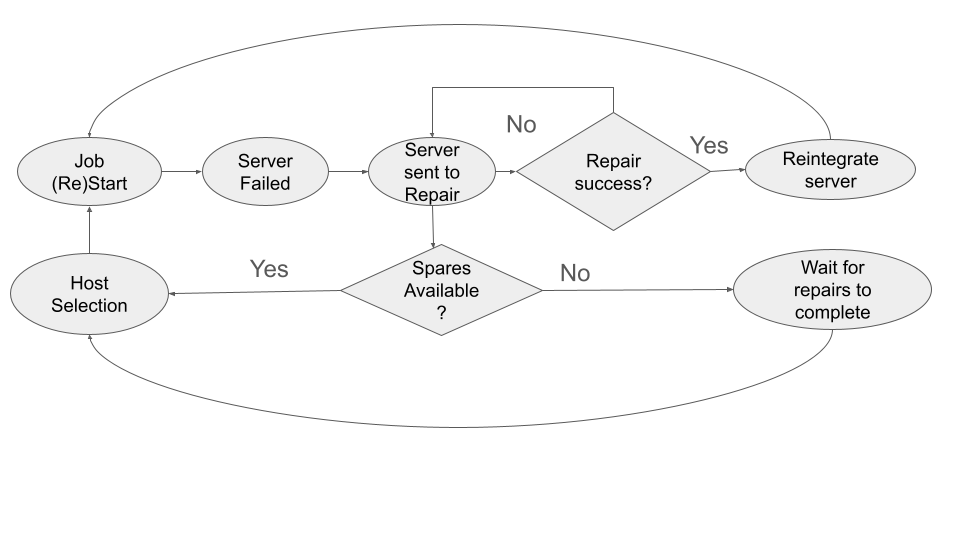}
    \vspace{-15mm}
    \caption{Overview of an AI job's scheduling by \relsim. }
    \label{fig:overview}
\end{figure}

Figure~\ref{fig:overview} shows an overview of the high-level algorithm as a flowchart. It shows the steps taken by \relsim to schedule jobs on the servers in the working and spare pools, and also the different situations that can arise. 

\subsection{Implementation}

We implemented \relsim in Python using the SimPy~\cite{scherfke2021simpy} simulation engine. Our implementation consists of about 2700 lines of Python code, including comments. \relsim is modular and easily extensible, with the developer only needing to extend classes to modify the simulator's behavior~\footnote{\rev{\relsim is publicly available at \url{https://github.com/karthikp-ubc/airesim} - this version was independently developed outside Meta based on this paper.}}

We describe the user's perspective instead of the developer's perspective. 
To use \relsim, the user provides two files. 
\begin{enumerate}
    \item Parameters (Params data class): This consists of all the parameters used for the simulation - these are the default values used when doing parameter sweeps. These can be constants, variables, or Python expressions. 
    \item Experiments (config.yaml / run.py): This is the main driver for the simulation. Each simulation run is an experiment that performs a parameter sweep (or two parameter sweeps), and collects statistics. Both the experiments and the statistics are configurable in this file. As explained below, there are two kinds of experiments: for one-way parameter sweeps and two-way sweeps. We can also configure the statistics collected in this file.
\end{enumerate}

To launch a simulation experiment, the user needs to instantiate either a one-way parameter sweep or a two-way parameter sweep. The former supports varying the values of one of the parameters, while the latter supports varying the values of two parameters simultanously. For example, to vary the parameter ``systematic failure fraction'', one would instantiate the class OneWaySweep as follows: OneWaySweep("Systematic Failure Fraction", "systematic\_failure\_fraction", [0.1, 0.2, 0.3 ]). The statistics for each experiment are written to a separate file. 

\section{Evaluation}
\label{sec:evaluation}

To evaluate \relsim, we perform different parameter sweeps for the various simulation parameters. The common element in all these simulations is that we vary the capacity of the working pool systematically (for four different values), along with one other parameter, for a two-way sweep. Table~\ref{tbl:params} shows the values of the different parameters we considered, for each value of the working pool size (note that these parameter values are hypothetical and not based on any observations of Meta's infrastructure). The goal of this simulation was to determine the size of the working pool at which the given training time is minimized. We plotted the total training time for different parameter values for different values of the working pool size, and hence lower values are better. 

\rev{Note that the failure rates used are consistent with publicly reported failures for LLama3 jobs~\cite{grattafiori2024llama}. We have also validated the results of \relsim using internal failure data at Meta. }

We considered values of \rev{4112,} 4128, 4160 and 4192 for the working pool size - so as to fit the minimum requirements of simulating a 4096 node training job, with \rev{16} warm standby servers. The above numbers correspond to a working pool capacity that is \rev{16, 32, 64 and 96} servers above the minimum number of servers needed for running the job, respectively. 

\begin{table*}
\centering
\begin{footnotesize}
\begin{tabular}{|c|c|c|}
\hline
     \textbf{Parameter} &  
     \textbf{Default Value} & 
     \textbf{Value Range Considered}
     \\
     \hline

     Random Failure Rate & $0.01/(24*60)$ & \{ $0.005/(24*60)$, $0.01/(24*60)$, $0.025/(24*60)$, $0.05/(24*60)$ \}  \\
     Systematic Failure Rate & 5 * random failure rate & \{3, 5, 10\} * random failure rate\\
     Systematic Failure Fraction & 0.15 & \{0.1, 0.15, 0.2\} \\
     Recovery Time (mins) & 20 & \{10, 20, 30\} \\
     Warm Standbys & 16 & \{4, 8, 16, 32\} \\
     Host Selection Time (mins) & 3 & \{1, 3, 5, 10\} \\
     Waiting Time (mins) & 20 & \{10, 20, 30\} \\ 
     Automated repair probability & 0.80 & \{0.70, 0.80, 0.90\} \\
     Auto repair failure probability & 0.40 & \{0.2, 0.4, 0.6\} \\
     Manual repair failure probability & 0.20 & \{ 0.1, 0.2, 0.3\} \\
     Auto repair time (mins) & 120 & \{60, 120, 180\} \\
     Manual repair time (mins) & $2 * 1440$ & \{ 1440, $2 * 1440$, $3 * 1440$ \}\\
     Working Pool Size & 4160 & \{ 4112, 4128, 4160, 4192 \}\\
     Spare Pool Size & 200 & \{200, 300, 400 \} \\
     \rev{Diagnosis probability} & \rev{0.8} & \rev{\{0.6, 0.8, 1.0\}} \\
     \hline
\end{tabular}
\end{footnotesize}
\caption{Parameters we used in the simulation and their values. The default value is the one we use when varying other parameters, and the value range shows the range we use when varying this parameter (one or two parameters at a time). }
\label{tbl:params}
\end{table*}

\begin{figure}
(a) \includegraphics[width=0.9\linewidth]{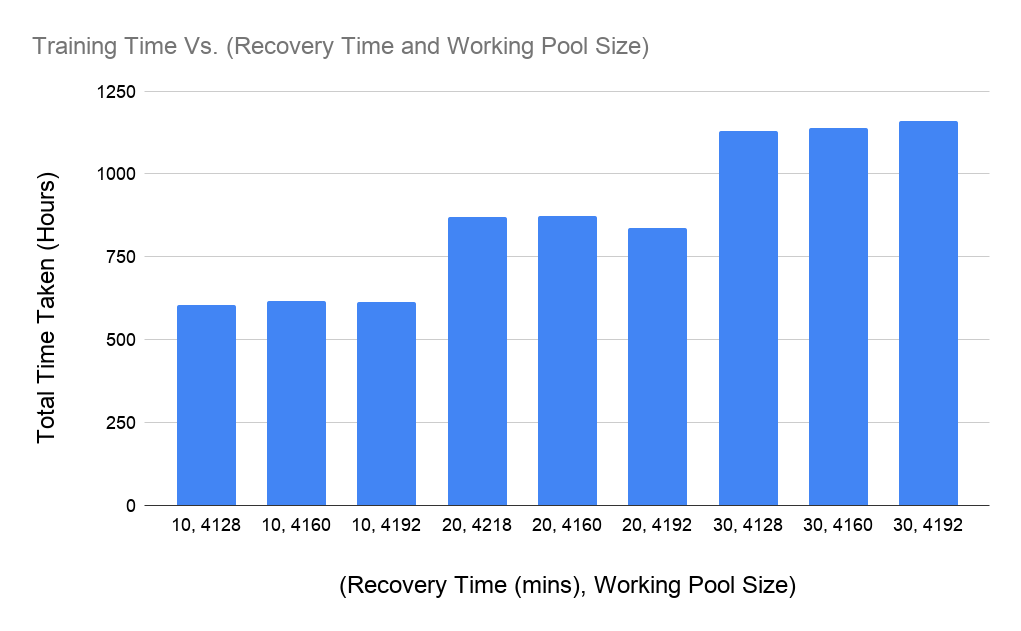} 
(b) \includegraphics[width=1.0\linewidth]{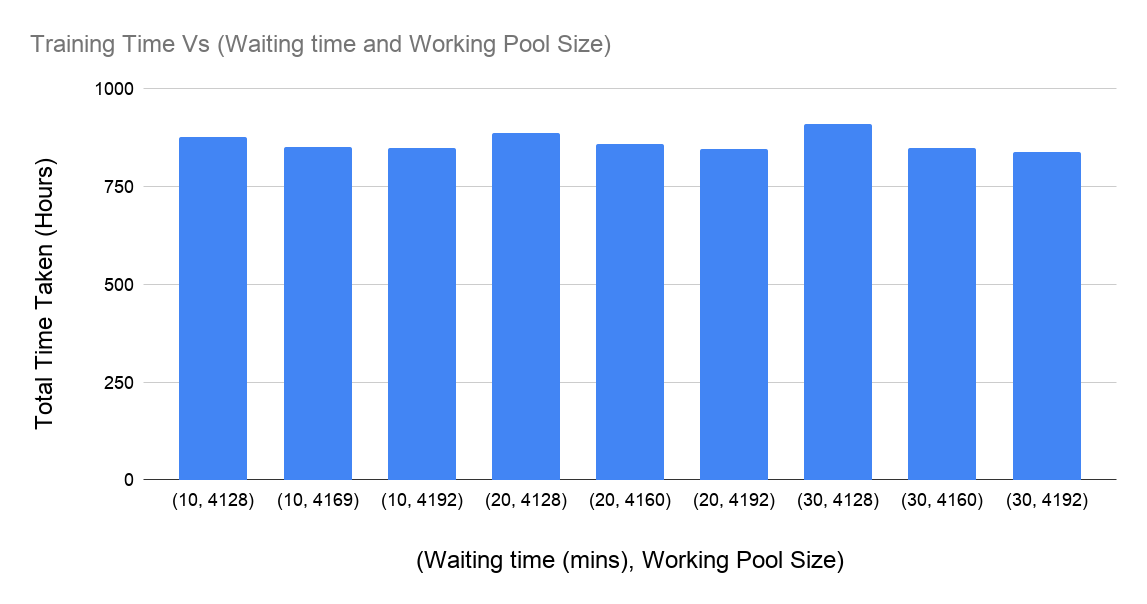} 
\caption{Graphs of the total Training Time in hours Vs. (a) Recovery time, in minutes, and (b) Waiting time, in minutes.}
\label{fig:graphs}
\end{figure}

For the simulation values in Table~\ref{tbl:params}, we find that \emph{none of the parameters has a significant impact on the training time except for the failure recovery time, and the waiting time for preempting a job from the spare pool.} We show the graphs for the recovery time in Figure~\ref{fig:graphs}(a), and the graph  for the waiting time in Figure~\ref{fig:graphs}(b). We do not show the graphs for the other parameters as they do not exhibit much variation.

We find that as the recovery time (Figure~\ref{fig:graphs}a) increases, the overall training time increases - this is because more time is spent on recovery than on the computation, which means it takes longer for the job to complete. However, for a given recovery time, there is a slight decrease in the training time as the number of servers in the working pool increases, due to the increase in the working pool's capacity. 

Similarly, as the waiting time (i.e., time for job preemption) increases, the overall time increases, though the increase is not as pronounced as that for the recovery time. This is again to be expected as the longer it takes to preempt a job from the spare pool, the longer the AI job stalls waiting for a server, and hence the longer is the training time. This increase is particularly seen when the working pool capacity is 0 servers over the minimum number of servers needed, as it means we need to preempt a job to move a server from the spare pool to the working pool whenever the job runs out of warm standbys. Overall, we find that (not surprisingly) having a small number of additional servers in the working pool is better than having none, but that a small number of additional servers (32) will suffice for the simulated system. \emph{Thus, based on these sets of parameters (Table~\ref{tbl:params}), there is no need to have more than 32 additional servers in the working pool, for a total working pool size of 4160.} This illustrates the value of \relsim as it allows optimizing capacity and avoiding wasteful expenditures. 

It is also interesting that most of the other parameters in the system did not have a substantial effect on the overall training time of the job (contrary to our expectations). This is likely because this simulation run/system is over-provisioned and hence has sufficient redundancy, to tolerate a relatively small number of failures. Also, the repair times we simulated are fairly low, and they allow the servers to be returned to the working pool after repair (with a high probability of successful repair) within a relatively few iterations. This obviates the need for more expensive mechanisms such as increasing the repair's efficacy (e.g., running more tests) or permanently removing servers that exceed a certain number of failures within a certain time window. However, we note that these results may not hold for other system configurations or failure rates. 
\section{Conclusion}
\label{sec:conclusion}

We presented \relsim, a discrete event simulator for understanding the effects of different parameters and tradeoffs in large-scale AI training jobs. Because such jobs are subject to both random and systematic failures, it becomes important to systematically reason about the reliability mechanisms in the cluster, and how they interact to tune them for optimal efficiency. We consider the different tradeoffs that arise in capacity planning for spares in the cluster to deal with systematic failures, and how much to provision for a given cluster configuration using \relsim. We find that \relsim can identify the optimal capacity for reliability, and hence avoid wasteful excess capacity. It can also help identify the important parameters in the system that should be optimized. 

\bibliographystyle{IEEEtran}
\bibliography{references}

@misc{chatgpt2025,
  author = {OpenAI},
  title = {{ChatGPT}},
  year = {2025},
  howpublished = {\url{https://chat.openai.com/}},
  note = {Accessed: 2025-01-23},
  version = {GPT-4o}
}

@book{mishra2023artificial,
  title={Artificial intelligence and hardware accelerators},
  author={Mishra, Ashutosh and Cha, Jaekwang and Park, Hyunbin and Kim, Shiho},
  year={2023},
  publisher={Springer}
}

@inproceedings{kokolis2025revisiting,
  title={Revisiting reliability in large-scale machine learning research clusters},
  author={Kokolis, Apostolos and Kuchnik, Michael and Hoffman, John and Kumar, Adithya and Malani, Parth and Ma, Faye and DeVito, Zachary and Sengupta, Shubho and Saladi, Kalyan and Wu, Carole-Jean},
  booktitle={2025 IEEE International Symposium on High Performance Computer Architecture (HPCA)},
  pages={1259--1274},
  year={2025},
  organization={IEEE}
}

@article{grattafiori2024llama,
  title={The llama 3 herd of models},
  author={Grattafiori, Aaron and Dubey, Abhimanyu and Jauhri, Abhinav and Pandey, Abhinav and Kadian, Abhishek and Al-Dahle, Ahmad and Letman, Aiesha and Mathur, Akhil and Schelten, Alan and Vaughan, Alex and others},
  journal={arXiv preprint arXiv:2407.21783},
  year={2024}
}

@inproceedings{Cunchi-2025,
author = {Lv, Cunchi and Shi, Xiao and Liang, Dong and Tan, Wenting and Zhao, Xiaofang},
title = {SpecInF: Exploiting Idle GPU Resources in Distributed DL Training via Speculative Inference Filling},
year = {2025},
isbn = {978-981-96-2829-2},
publisher = {Springer-Verlag},
address = {Berlin, Heidelberg},
url = {https://doi.org/10.1007/978-981-96-2830-8_12},
doi = {10.1007/978-981-96-2830-8_12},
booktitle = {Network and Parallel Computing: 20th IFIP WG 10.3 International Conference, NPC 2024, Haikou, China, December 7–8, 2024, Proceedings, Part I},
pages = {146–158},
numpages = {13},
location = {Haikou, China}
}

@ARTICLE{Mitra-2025,
  author={Mitra, Subhasish and Banerjee, Subho S. and Dixon, Martin and Fuller, Mike and Govindaraju, Rama and Hochschild, Peter and Liu, Eric X. and Parthasarathy, Bharath and Ranganathan, Parthasarathy},
  journal={IEEE Design and Test}, 
  title={Silent Data Corruption by 10× Test Escapes Threatens Reliable Computing}, 
  year={2025},
  volume={42},
  number={6},
  pages={40-53},
  doi={10.1109/MDAT.2025.3602741}}

@article{dixit2021silent,
  title={Silent data corruptions at scale},
  author={Dixit, Harish Dattatraya and Pendharkar, Sneha and Beadon, Matt and Mason, Chris and Chakravarthy, Tejasvi and Muthiah, Bharath and Sankar, Sriram},
  journal={arXiv preprint arXiv:2102.11245},
  year={2021}
}

@INPROCEEDINGS{jiao-2025,
  author={Jiao, Xun and Pandey, Abhinav and Pattabiraman, Karthik and Lin, Fred},
  booktitle={2025 55th Annual IEEE/IFIP International Conference on Dependable Systems and Networks - Supplemental Volume (DSN-S)}, 
  title={Large-Scale AI Infra Reliability: Challenges, Strategies, and Llama 3 Training Experience}, 
  year={2025},
  volume={},
  number={},
  pages={140-146},
  doi={10.1109/DSN-S65789.2025.00054}}

@book{trivedi2001probability,
  title={Probability and statistics with reliability, queuing, and computer science applications},
  author={Trivedi, Kishor S},
  year={2001},
  publisher={John Wiley \& Sons}
}

@book{fishman2001discrete,
  title={Discrete-event simulation: modeling, programming, and analysis},
  author={Fishman, George S},
  volume={537},
  year={2001},
  publisher={Springer}
}

@inproceedings{hwremediationatscale,
  author={Lin, Fan and Beadon, Matt and Dixit, Harish Dattatraya and Vunnam, Gautham and Desai, Amol and Sankar, Sriram},
  booktitle={2018 48th Annual IEEE/IFIP International Conference on Dependable Systems and Networks Workshops (DSN-W)}, 
  title={Hardware Remediation at Scale}, 
  year={2018},
  volume={},
  number={},
  pages={14-17},
  doi={10.1109/DSN-W.2018.00015}}

@article{elnozahy2004checkpointing,
  title={Checkpointing for peta-scale systems: A look into the future of practical rollback-recovery},
  author={Elnozahy, Elmootazbellah N and Plank, James S},
  journal={IEEE Transactions on Dependable and Secure Computing},
  volume={1},
  number={2},
  pages={97--108},
  year={2004},
  publisher={IEEE}
}

@misc{dixit2022detectingsilentdatacorruptions,
      title={Detecting silent data corruptions in the wild}, 
      author={Harish Dattatraya Dixit and Laura Boyle and Gautham Vunnam and Sneha Pendharkar and Matt Beadon and Sriram Sankar},
      year={2022},
      eprint={2203.08989},
      archivePrefix={arXiv},
      primaryClass={cs.AR},
      url={https://arxiv.org/abs/2203.08989}, 
}

@article{quin2024b,
  title={A/B testing: A systematic literature review},
  author={Quin, Federico and Weyns, Danny and Galster, Matthias and Silva, Camila Costa},
  journal={Journal of Systems and Software},
  volume={211},
  pages={112011},
  year={2024},
  publisher={Elsevier}
}

@article{scherfke2021simpy,
  title={SimPy},
  author={Scherfke, Stefan and L{\"u}nsdorf, Ontje and Grayson, Peter and LaFevers, Eric and Pinckney, Thomas and Klein, Cristian and Vaidya, Sundar and Reis, Larissa and Reed, Sean and Liu, Zhe and others},
  journal={URL https://github. com/simpx/simpy},
  year={2021}
}

@inbook{Jiang-2025,
author = {Jiang, Zhihan and Huang, Junjie and Yu, Guangba and Chen, Zhuangbin and Li, Yichen and Zhong, Renyi and Feng, Cong and Yang, Yongqiang and Yang, Zengyin and Lyu, Michael},
title = {L4: Diagnosing Large-scale LLM Training Failures via Automated Log Analysis},
year = {2025},
isbn = {9798400712760},
publisher = {Association for Computing Machinery},
address = {New York, NY, USA},
url = {https://doi.org/10.1145/3696630.3728531},
booktitle = {Proceedings of the 33rd ACM International Conference on the Foundations of Software Engineering},
pages = {51–63},
numpages = {13}
}

@misc{simgrid,
      title={SimGrid: a Sustained Effort for the Versatile Simulation of Large Scale Distributed Systems}, 
      author={Henri Casanova and Arnaud Giersch and Arnaud Legrand and Martin Quinson and Frédéric Suter},
      year={2013},
      eprint={1309.1630},
      archivePrefix={arXiv},
      primaryClass={cs.DC},
      url={https://arxiv.org/abs/1309.1630}, 
}

@article{SHARPE,
author = {Trivedi, Kisho S. and Sahner, Robin},
title = {SHARPE at the age of twenty two},
year = {2009},
issue_date = {March 2009},
publisher = {Association for Computing Machinery},
address = {New York, NY, USA},
volume = {36},
number = {4},
issn = {0163-5999},
url = {https://doi.org/10.1145/1530873.1530884},
doi = {10.1145/1530873.1530884},
journal = {SIGMETRICS Perform. Eval. Rev.},
month = mar,
pages = {52–57},
numpages = {6}
}
\end{document}